\documentclass[pre,twocolumn,superscriptaddress]{revtex4-1}

\usepackage{dcolumn}
\usepackage{amsmath}
\usepackage{graphicx}
\usepackage{calrsfs}
\usepackage{bm}
\usepackage[table,xcdraw]{xcolor}
\usepackage{enumitem}
\usepackage{hyperref}
\usepackage{cancel}

\newcommand{\dd}{\mathrm{d}}
\newcommand{\Ord}{\mathrm{O}}

\renewcommand{\vec}{\bm}
\newcommand\Beta{\mathrm{B}}
\newcommand\etal{\textit{et~al.}}

\begin{document}
\title{Mutual information and the encoding of contingency tables}

\author{Maximilian Jerdee}
\affiliation{Department of Physics, University of Michigan, Ann Arbor, Michigan 48109, USA}

\author{Alec Kirkley}
\affiliation{Institute of Data Science, University of Hong Kong, Hong Kong}
\affiliation{Department of Urban Planning and Design, University of Hong Kong, Hong Kong}
\affiliation{Urban Systems Institute, University of Hong Kong, Hong Kong}

\author{M. E. J. Newman}
\affiliation{Department of Physics, University of Michigan, Ann Arbor, Michigan 48109, USA}
\affiliation{Center for the Study of Complex Systems, University of Michigan, Ann Arbor, Michigan 48109, USA}

\begin{abstract}
Mutual information is commonly used as a measure of similarity between competing labelings of a given set of objects, for example to quantify performance in classification and community detection tasks.  As argued recently, however, the mutual information as conventionally defined can return biased results because it neglects the information cost of the so-called contingency table, a crucial component of the similarity calculation.  In principle the bias can be rectified by subtracting the appropriate information cost, leading to the modified measure known as the reduced mutual information, but in practice one can only ever compute an upper bound on this information cost, and the value of the reduced mutual information depends crucially on how good a bound is established.  In this paper we describe an improved method for encoding contingency tables that gives a substantially better bound in typical use cases, and approaches the ideal value in the common case where the labelings are closely similar, as we demonstrate with extensive numerical results.
\end{abstract}
\maketitle

%%%%%%%%%%%%%%%%%%%%%%%%%%%%%%%%%%%%%%%%%%%%%%%%%%%%%%%%%%%%%%%%%%%%%%%%%%%%%%%%%%%%%%%%%%%%%%%%%%%%%%%%%%%%%%%%%%%%%%%%%%%%%%%%%%%
\section{Introduction}
\label{sec:intro}
A common task in data analysis is the comparison of two different labelings of the same set of objects.  How well do demographics predict political affiliation?  How accurately do blood tests predict clinical outcomes?  How well do clustering algorithms recover known classes of items?  Questions like these, in which an experimental or computational labeling is compared against a ``ground truth,'' are commonly addressed using the information theoretic measure known as mutual information~\cite{CT06}.

Mutual information is a measure of how easy it is to describe one labeling of a set of objects if we already know another labeling.  Specifically, it measures how much less information it takes (in the Shannon sense) to communicate the first labeling if we know the second versus if we do not.  As an example, mutual information is commonly used in network science to evaluate the performance of algorithms for network community detection~\cite{DDDA05}.  One takes a network whose community structure is already known and applies a community detection algorithm to it to infer the communities.  Then one uses mutual information to compare the output of the algorithm to the ground truth node labels.  Algorithms that consistently achieve high mutual information scores are considered good.

Mutual information has a number of appealing properties as a tool for comparing labelings.  It is invariant under permutations of the labels in either labeling, so that labelings do not have to be aligned before comparison.  It also generalizes gracefully to the case where the number of distinct labels is not the same in the two labelings.  On the other hand, the mutual information in its most common form also has some significant drawbacks and, in particular, it is known to be biased towards labelings with a large number of distinct labels.  Various proposals have been made for mitigating this effect~\cite{Dom02,VEB10,Zhang15,AP17,NCY20}.  In this paper we focus on the recently proposed \textit{reduced mutual information}~\cite{NCY20}, which improves on the standard measure by carefully accounting for subleading terms in the information that are normally neglected.

Any version of the mutual information is an approximation to the true information cost of the labelings being compared.  One computes the information cost by defining some encoding scheme for labelings and then counting the number of bits needed to specify a labeling within that encoding.  In this paper we highlight two common pitfalls that occur when quantifying information cost in this way, which produce errors in opposite directions.  The first is the neglect of the information cost of certain parts of the transmission process, resulting in an underestimate of the total cost of transmission.  The standard, unreduced mutual information is an example: it does not include the cost to transmit the ``contingency table'' that summarizes the relationship between the two labelings, causing it to underestimate---sometimes drastically---the total information cost, particularly for labelings with many groups.

The second pitfall, and the focus of this paper, is the use of inefficient encoding schemes, which result in overestimates of information cost.  The reduced mutual information, in its conventional form, suffers from this issue because it uses a relatively crude encoding of the contingency table.  In this paper we offer an improved encoding that gives better bounds on the value of the reduced mutual information, different enough to change outcomes in some practical situations, as we demonstrate with a selection of illustrative examples. Code implementing our new similarity measure may be found at \url{https://github.com/maxjerdee/reduced_mutual_information}.

%%%%%%%%%%%%%%%%%%%%%%%%%%%%%%%%%%%%%%%%%%%%%%%%%%%%%%%%%%%%%%%%%%%%%%%%%%%%%%%%%%%%%%%%%%%%%%%%%%%%%%%%%%%%%%%%%%%%%%%%%%%%%%%%%%%
\section{Conditional entropy and\\mutual information}
\label{sec:mutual-information}
To motivate our discussion, we first rederive the conventional mutual information using the language of information transmission, before progressing to the reduced mutual information and its variants.

\subsection{Mutual information}
\label{sec:unreduced-mutual-information}
Mutual information can be thought of in terms of the amount of information it takes to transmit a labeling from one person to another.  Suppose, first, that we want to transmit to a receiver a discrete quantity~$X$, which can take any one of a known finite set of~$N$ values.  For example, we could be transmitting the outcome of a coin flip~$X \in \{\text{heads},\text{tails}\}$ or one possible labeling of a group of objects.  If we assign to each possible value of~$X$ a unique binary string, we can convey any particular value by transmitting the appropriate string.  The minimum length of string that can encode all $N$ values is
\begin{equation}
H = \bigl\lceil \log_2 N\bigr\rceil \simeq \log_2 N,
\label{eq:flat-information}
\end{equation}
where $\lceil x\rceil$ denotes the smallest integer not less than~$x$.  This tells us the number of bits of information needed to transmit~$X$.

Conventionally one uses base-2 logarithms in equations like~\eqref{eq:flat-information}, which gives~$H$ in units of bits, as here.  Some authors use natural logarithms, which changes the results by a constant factor, but the difference is not an important one.  In this paper we use base-2 logarithms for explicit numerical calculations, but our formal results are all independent of the base of the logarithm and one can use any base one wishes provided one is consistent.  Henceforth, we will write logarithms without any base indicated.

Suppose now that~$X$ is actually a labeling~$g$ of a set of~$n$ objects, with each object having exactly one label and each label having the same~$q_g$ possible values, which we represent by integers in the range~$1\ldots q_g$.  Then there are~$N = q_g^n$ possible values of the entire labeling and hence any labeling can be transmitted using an amount of information
\begin{equation}
H(g) = \log N = n \log q_g.
\label{eq:flat-labeling}
\end{equation}

This, however, is not necessarily the most efficient way to transmit such a labeling.  In particular, if different labels occur with different frequencies then a more efficient encoding may be possible, resulting in a smaller information cost.  The standard encoding used to do this has three parts.  First we transmit the number of groups~$q_g$ in the labeling.  The maximum possible value of~$q_g$ is~$n$, so if we use a simple ``flat'' encoding as in Eq.~\eqref{eq:flat-information}, then transmitting any particular value requires information~$H(q_g) = \log n$.  Next we transmit a vector~$n^{(g)}$ of~$q_g$ integers~$n^{(g)}_r$ equal to the number of objects having each label~$r$.  By definition, the~$n^{(g)}_r$ sum to~$n$, and the number of ways to choose $q_g$ nonnegative integers that sum to~$n$ is~$\binom{n+q_g-1}{q_g-1}$, so if we again use a flat encoding to transmit~$n^{(g)}$ the information cost will be
\begin{equation}
H(n^{(g)}|q_g) = \log \binom{n+q_g-1}{q_g-1}.
\end{equation}

Finally, we transmit the labeling~$g$ itself, choosing only from among those that have the correct multiplicities~$n^{(g)}_r$ of the labels.  The number of such labelings is given by the multinomial~$n!/\prod_r n^{(g)}_r!$ and hence, choosing a flat encoding one more time, the information cost is
\begin{equation}
H(g|n^{(g)}) = \log {n!\over\prod_r n^{(g)}_r!}.
\label{eq:Hgng}
\end{equation}
Putting everything together, the complete cost to transmit the labeling is
\begin{align}
H(g) &= H(q_g) + H(n^{(g)}|q_g) + H(g|n^{(g)}) \nonumber\\
  &= \log n + \log \binom{n+q_g-1}{q_g-1} + \log {n!\over\prod_r n^{(g)}_r!}.
\label{eq:HRg}
\end{align}
This three-step scheme is not the only one that could be applied to this problem, but it is a fairly efficient one in the common case of a small number of groups~$q_g\ll n$ with potentially unequal sizes, and it is the one on which the conventional definition of labeling entropy is based.

The conventional definition, however, ignores all but the last term and approximates the entropy as
\begin{align}
H_0(g) &= \log {n!\over\prod_r n^{(g)}_r!}\,.\label{eq:H0g}
\end{align}
In most cases this is a good approximation.  The other terms are subleading contributions---they grow more slowly with~$n$ than the leading term---and in practice they are negligible even for quite modest values of~$n$.  Commonly one also makes a further approximation, applying Stirling's formula to each of the factorials, which gives the Shannon form of the entropy~$H_0(g) = -n \sum_r p_r \log p_r$, where~$p_r = n_r^{(g)}/n$ is the probability that a randomly chosen object has label~$r$.

Now consider the corresponding encoding scheme for mutual information.  Suppose that we have two different labelings of the same set of objects, a ground-truth labeling~$g$ which represents the ``true'' labels, and a candidate labeling~$c$, generated for instance by some sort of algorithm.  The mutual information~$I(g;c)$ between the two is the amount of information saved when transmitting the truth~$g$ if the receiver already knows the candidate~$c$.  We can write this information as the total information or entropy needed to transmit~$g$ on its own, minus the conditional entropy, the amount to transmit~$g$ given prior knowledge of~$c$:
\begin{equation}
I(g;c) = H(g) - H(g|c).
\label{eq:I-Hg1-minus}
\end{equation}

\begin{figure}
\centering
\includegraphics[width=0.9\linewidth]{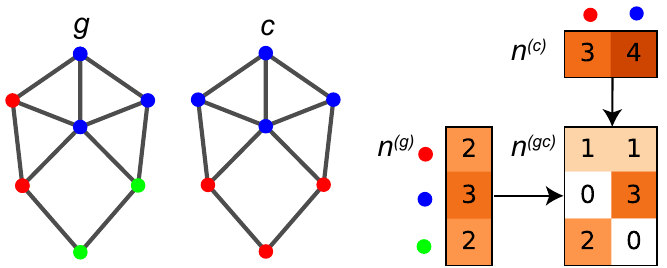}
\caption{A contingency table for two labelings represented as colorings of the nodes of a network, one with three colors and one with two.  The entries in the~$3\times2$ contingency table~$n^{(gc)}$ count the number of nodes that have each combination of labels.  The row and column sums~$n^{(g)}$ and~$n^{(c)}$ count the number of nodes with each label in the two labelings.  Note that, although we illustrate the contingency table with an application to a network, the table itself is independent of the network structure.}
\label{fig:Contingency-Example}
\end{figure}

For the first term, we use the three-part encoding scheme described above, Eq.~\eqref{eq:HRg}.  For the second we use a similar multi\-part scheme, but one that now takes advantage of the receiver's knowledge of~$c$.  In this scheme we first communicate~$q_g$ and~$n^{(g)}$ as before, at the same information cost of~$H(q_g) + H(n^{(g)}|q_g)$.  Then we communicate a \textit{contingency table}~$n^{(gc)}$, a matrix with elements~$n_{rs}^{(gc)}$ equal to the number of objects that simultaneously belong to group~$r$ in the ground truth~$g$ and group~$s$ in the candidate labeling~$c$.  Figure~\ref{fig:Contingency-Example} shows an example of a contingency table for two labelings of the nodes in a small network.

The contingency table consists of non-negative integer elements and its row and column sums are equal to the multiplicities of the labels in~$g$ and~$c$ respectively:
\begin{equation}
\sum_{s=1}^{q_c} n_{rs}^{(gc)} = n_r^{(g)}, \qquad \sum_{r=1}^{q_g} n_{rs}^{(gc)} = n_s^{(c)}.
\label{eq:n12Constraints}
\end{equation}
Since the receiver already knows the values of~$n^{(g)}$ and~$n^{(c)}$ (the former because we just transmitted it and the latter because they know~$c$), only contingency tables with these row and column sums need be considered.  The information cost to transmit the contingency table with a flat encoding is then equal to~$\log\Omega(n^{(g)},n^{(c)})$, where~$\Omega(n^{(g)},n^{(c)})$ is the number of possible tables with the required row and column sums.  There is no known general expression for this number, but approximations exist that are good enough for practical purposes~\cite{DE85,BH10,JKN24}.

Finally, having transmitted the contingency table, it remains only to transmit the ground-truth labeling itself, where we need consider only those labelings consistent with the contingency table and the candidate labeling~$c$.  The number of such labelings is~$\prod_s n_s^{(c)}!/\prod_{rs} n_{rs}^{(gc)}!$, so the information needed to uniquely identify one of them~is
\begin{equation}
H(g|c,n^{(gc)
}) = \log \frac{\prod_s n_s^{(c)}!}{\prod_{rs} n_{rs}^{(gc)}!}.
\end{equation}
Putting everything together, the total conditional information is then
\begin{align}
H(g|c) &= H(q_g) + H(n^{(g)}|q_g) \nonumber\\
    &\qquad{} + H(n^{(gc)}|n^{(g)},n^{(c)}) + H(g|c,n^{(gc)}) \nonumber\\
    &= \log n + \log \binom{n + q_g - 1}{q_g - 1} \nonumber\\
    &\qquad{} + \log \Omega(n^{(g)},n^{(c)})
     + \log \frac{\prod_s n_s^{(c)}!}{\prod_{rs} n_{rs}^{(gc)}!}.
\label{eq:HRgGc}
\end{align}
In typical applications the number of labelings compatible with the contingency table is much smaller than the total number of labelings, and hence the amount of information needed to transmit the ground truth using this encoding is substantially smaller than Eq.~\eqref{eq:HRg}.  The difference---the amount saved---is the quantity we call the mutual information:
\begin{align}
I(g;c) &= H(g) - H(g|c) \nonumber\\
  &= H(q_g) + H(n^{(g)}|q_g) + H(g|n^{(g)})
  \nonumber\\
  &\qquad - \big[ H(q_g) + H(n^{(g)}|q_g) \nonumber\\
    &\qquad\qquad + H(n^{(gc)}|n^{(g)},n^{(c)}) + H(g|c,n^{(gc)}) \bigr] \nonumber\\
  &= H(g|n^{(g)}) - H(g|c,n^{(gc)}) - H(n^{(gc)}|n^{(g)},n^{(c)}) \nonumber\\
  &= \log \frac{n!\prod_{rs} n_{rs}^{(gc)}!}{\prod_r n_r^{(g)}! \prod_s n_s^{(c)}!} - \log \Omega(n^{(g)},n^{(c)}).
\label{eq:RMI}
\end{align}

Once again this encoding is not necessarily the most efficient one, but it works well in practical situations and forms the basis for the conventional definition of mutual information.  And once again the conventional definition drops the subleading term, retaining only the first term in~\eqref{eq:RMI}:
\begin{equation}
I_0(g;c) =\log \frac{n!\prod_{rs} n_{rs}^{(gc)}!}{\prod_r n_r^{(g)}! \prod_s n_s^{(c)}!}.
\label{eq:I0}
\end{equation}
Commonly one again also applies Stirling's approximation, which leads to the familiar expression for the mutual information $I_0(g;c) = n \sum_{rs} p_{rs} \log (p_{rs}/p_r p_s)$, where $p_{rs} = n_{rs}^{(gc)}/n$ is the joint probability that a randomly chosen object has label $r$ in the ground truth and $s$ in the candidate labeling.

Though the principles behind them are similar, an important practical difference between Eq.~\eqref{eq:I0} for the mutual information and Eq.~\eqref{eq:H0g} for the entropy is that the subleading term neglected in the mutual information is typically larger and can significantly affect the overall value.  It is the neglect of this term that produces the bias towards an excessive number of groups described in the introduction.  The cure for this bias is to retain the subleading term, which leads to the measure known as the reduced mutual information.

%%%%%%%%%%%%%%%%%%%%%%%%%%%%%%%%%
\subsection{Reduced mutual information}
\label{sec:reduced-mutual-information}
Equation~\eqref{eq:I0} defines the standard mutual information~$I_0$, which neglects subleading behavior.  In the limit of large~$n$ this is a good approximation, but for finite~$n$, including values large enough to be of practical consequence, the subleading term can contribute significantly.  In this section, we demonstrate how this gives rise to a bias in favor of labelings with larger numbers of groups and how simply retaining the subleading term removes this bias.

The full expression in Eq.~\eqref{eq:RMI} is known as the reduced mutual information, with this particular version (we will shortly consider others) distinguished by the fact that it assumes a flat encoding when transmitting the contingency table.  We will denote this measure by~$I_{\text{flat}}$:
\begin{equation}
I_{\text{flat}}(g;c) = \log \frac{n!\prod_{rs} n_{rs}^{(gc)}!}{\prod_r n_r^{(g)}! \prod_s n_s^{(c)}!} - \log \Omega(n^{(g)},n^{(c)}).
\label{eq:RMI2}
\end{equation}
The moniker ``reduced'' derives from the fact that the~$-\log\Omega$ term is always negative and so reduces the value of the mutual information relative to the conventional definition of Eq.~\eqref{eq:I0}, but we emphasize that functionally we are simply retaining terms that are usually neglected.  As mentioned previously, there is no general closed-form expression for the number~$\Omega(n^{(g)},n^{(c)})$ of contingency tables with given row and column sums, and its numerical computation is \#P-hard in general~\cite{DKM97} and hence intractable for all but the smallest of examples.  In practice, therefore, the value must be approximated.  In this paper we make use of the ``effective columns'' approximation of~\cite{JKN24}, which has good performance over a wide range of situations and a simple closed-form expression:
\begin{align}
&\Omega(n^{(g)},n^{(c)}) \simeq \nonumber\\
  &\binom{n+q_c \alpha - 1}{q_c \alpha - 1}^{-1} \prod_{s=1}^{q_c}\! \binom{n_s^{(c)} + \alpha - 1}{ \alpha - 1} \prod_{r=1}^{q_g}\! \binom{n_r^{(g)}+q_c-1}{q_c - 1},
\label{eq:omega-approx}
\end{align}
where
\begin{equation}
\alpha = \frac{n^2 - n + \bigl(n^2 - R\bigr)/q_c}{R - n}, \quad
R = \sum_r (n_r^{(g)})^2.
\end{equation}
This estimate differs from the one originally used for the reduced mutual information in~\cite{NCY20}, but we favor it here since it performs better in certain regimes.

To understand the importance of the contingency table term in the mutual information, consider the simple case where the candidate labeling~$c$ places every object in a group of its own:~$c = (1,\ldots,n)$.  No matter what the ground truth labeling is, this choice of~$c$ clearly contains no information about it whatsoever, so we expect the mutual information to be zero.  But it is not.  We have~$n^{(c)}_s!=1$ for all~$s$ in this case, while the contingency table has a single~1 in each column and all other elements are~0, so~$n_{rs}!=1$ for all~$r,s$, and hence the conventional mutual information of Eq.~\eqref{eq:I0} simplifies to
\begin{equation}
I_0(g;c) = \log \frac{n!}{\prod_r n_r^{(g)}!} = H_0(g).
\end{equation}
This answer is as wrong as it possibly could be: we expect the mutual information to take the minimum value of zero, but instead it is equal to the entropy~$H_0(g)$, which is its maximum possible value, since the largest amount of information we can save by knowing~$c$ when we transmit~$g$ is equal to the entire information~$H_0(g)$.  In other words, the conventional mutual information would have us believe that this candidate labeling which puts every object in its own group tells us everything there is to know about the true labeling~$g$, when in fact it tells us nothing at all.

The reason for this dramatic failure is that in this case the contingency table itself uniquely defines~$g$, so neglecting it puts the mutual information in error by an amount equal to the complete information cost of the ground truth.  If we include the cost of transmitting the contingency table, this erroneous behavior disappears.  We can calculate the number $\Omega(n^{(g)},n^{(c)})$ of contingency tables exactly for this example.  Since there is just a single~1 in every column of the table, the number of tables is
\begin{equation}
\Omega(n^{(g)},n^{(c)}) = \frac{n!}{\prod_r n_r^{(g)}!},
\end{equation}
and the reduced mutual information is
\begin{equation}
I_{\text{flat}}(g;c) = I_0(g;c) - \log \frac{n!}{\prod_r n_r^{(g)}!} = 0,
\end{equation}
which is now the correct answer.

%%%%%%%%%%%%%%%%%%%%%%%%%%%%%%%%%
\subsection{Improved encodings}
\label{sec:improved-encoding}
The reduced mutual information offers a significant improvement over the traditional measure for finite-sized systems, particularly when the candidate labeling has a large number of distinct label values.  And, as we have seen, it gives exactly the correct answer in the case where every object is in a group of its own.  In this paper, however, we argue that the reduced mutual information, as it is usually defined, is itself an imperfect measure, and in particular that it often overcorrects for the flaws of traditional mutual information because the encoding scheme used for both~$H(g)$ and~$H(g|c)$ is inefficient and poorly approximates the best possible encoding.  All calculated entropies are merely upper bounds on the true value: calculating the information cost of transmitting a labeling using a specific encoding guarantees that no more than that amount of information is needed, but it is possible that a better encoding exists that can do the job with less.  In this section, we propose more efficient encodings that give better bounds on the entropy and the conditional entropy, particularly in the common case where the two partitions~$g$ and~$c$ are similar.

The central observation behind our proposed encodings is that quantities like~$n^{(g)}$ and~$n^{(gc)}$ usually have unevenly distributed elements, sometimes strongly so.  For example, mutual information is most often used to compare labelings that are quite similar, which means the elements of the contingency table are very non-uniform---those that correspond to common pairs of labels are large, while all the others are small.  This in turn means that choices of the contingency table with these properties are much more likely to occur than others and hence that a ``flat'' encoding that assumes all choices are equally likely is inefficient.  By using an encoding that allows for a non-uniform distribution, we can save a substantial amount of information and achieve a better approximation of the mutual information.

The encodings we propose are based on the symmetric Dirichlet-multinomial distribution, a standard, one-parameter family of discrete distributions over~$q$-vectors~$X$ of non-negative integer elements that sum to a given total~$N$. The distribution is derived from a two-part generative process in which, first, a set of~$q$ probabilities~$p_1\ldots p_q$ are drawn from a symmetric Dirchlet distribution with concentration parameter~$\alpha\ge0$, and then a set of~$q$ integers~$X_1\ldots X_q$ are drawn from a multinomial distribution with those probability parameters.  The resulting distribution over~$X$ is given by
\begin{align}
P(X|N,q,\alpha) &= \int \underbrace{{\prod_{r=1}^q p_r^{\alpha-1}\over\Beta(\alpha)}\vphantom{\prod_r}}_{\text{Dirichlet}} \>
  \underbrace{N! \prod_{r=1}^q {p_r^{X_r}\over X_r!}}_{\text{Multinomial}}
  \dd\vec{p},
\end{align}
where~$\Beta(\alpha)$ is the multivariate beta function and the integral is over the simplex of non-negative values~$p_r$ such that~$\sum_r p_r = 1$.  Performing the integral then gives the standard expression for the Dirichlet-multinomial distribution:
\begin{align}
    P(X|N,q,\alpha) = \binom{N + q \alpha - 1}{q \alpha - 1}^{-1} \prod_{r = 1}^q \binom{X_r + \alpha - 1}{\alpha - 1},
\label{eq:DM-dist}
\end{align}
where for non-integer~$\alpha$ we generalize the binomial coefficient in the obvious way:
\begin{equation}
\Bigl( {n\atop k} \Bigr) = {\frac{\Gamma(n+1)}{\Gamma(k+1)\Gamma(n-k+1)}},
\end{equation}
with~$\Gamma(x)$ being the standard gamma function. 

If~$\alpha = 1$, the Dirichlet-multinomial distribution is uniform over all vectors~$X$ of non-negative integers that sum to~$N$:
\begin{align}
    P(X|N,q,\alpha = 1) = \binom{N - q - 1}{q - 1}^{-1}.
\end{align}
Smaller values~$0\le \alpha < 1$ produce a distribution biased towards more heterogeneous~$X$.  In the extreme limit where~$\alpha \to 0$ (which we will denote as $\alpha = 0$) the distribution is supported only on vectors that have a single nonzero entry equal to~$N$:
\begin{align}
P(X|N,q,\alpha = 0) = \biggl\lbrace\begin{array}{ll}
  1/q & \mbox{if~$X$ has one nonzero entry,} \\
  0   & \mbox{otherwise.}
    \end{array}
\end{align}
Conversely, for~$\alpha>1$ the Dirichlet-multinomial distribution favors vectors~$X$ with more uniform entries, and in the limit~$\alpha \to \infty$ it approaches the symmetric multinomial distribution where $p_r = 1/q$ for all~$r$:
\begin{align}
P(X|N,q,\alpha \to \infty) = \frac{N!}{\prod_{r = 1}^q X_r!} (1/q)^N.
\end{align}
Different choices of the parameter~$\alpha$ thus place more or less weight on different types of vectors~$X$.  

We can use the Dirichlet-multinomial distribution to improve the encoding of the group sizes~$n^{(g)}$ and so better approximate the total unconditional information cost~$H(g)$.  The information cost used in the definition of the standard reduced mutual information is
\begin{align}
H_{\text{flat}}(g) &= H(q_g) + H(n^{(g)}|q_g) + H(g|n^{(g)}),
\end{align}
where as previously the subscript ``flat'' indicates the flat encoding that assumes equal probability for all outcomes at each step.  Here we propose an alternative approach that still uses flat encodings for~$q_g$ and~$g$ but uses a nonuniform Dirichlet-multinomial distribution for the group sizes~$n^{(g)}$.

Generally when transmitting a sequence of~$n$ values with unequal probabilities such that value~$r$ occurs~$n_r$ times, the information cost is given by Eq.~\eqref{eq:Hgng}:
\begin{align}
\log {n!\over \prod_r n_r!} &\simeq n\log n - n - \sum_r (n_r \log n_r - n_r)
  \nonumber\\
  &= - \sum_r n_r \log p_r,
\label{eq:sterling}
\end{align}
where~$p_r = n_r/n$ is the probability of value~$r$ and we have approximated the factorials using Stirling's formula.  Equation~\ref{eq:sterling} tells us that the information cost to transmit the value~$r$ is simply~$-\log p_r$. Applying this observation to the Dirichlet-multinomial distribution we can calculate the total information cost to transmit a vector~$n^{(g)}$ drawn from the Dirichlet-multinomial distribution with concentration parameter~$\alpha_g$:
\begin{align}
H(&n^{(g)}|q_g,\alpha_g) = -\log P(n^{(g)}|\alpha_g) \label{eq:HngGalphag}
  \nonumber\\
  &= \log \binom{n + q_g \alpha_g - 1}{q_g \alpha_g - 1} - \sum_{r=1}^{q_g} \log \binom{n_r^{(g)} + \alpha_g - 1}{\alpha_g - 1}.
\end{align}
The optimal encoding for transmitting~$n^{(g)}$ within the Dirichlet-multinomial family is given by the minimum of this expression with respect to~$\alpha_g$, which is also equivalent to simply maximizing~$P(n^{(g)}|\alpha_g)$, i.e.,~to finding the maximum-likelihood value of~$\alpha_g$.  In practice we can find the maximum-likelihood value with standard numerical optimization techniques, as described in Appendix~\ref{app:alpha-parameter}.

\begin{figure}
\centering
\includegraphics[width=\linewidth]{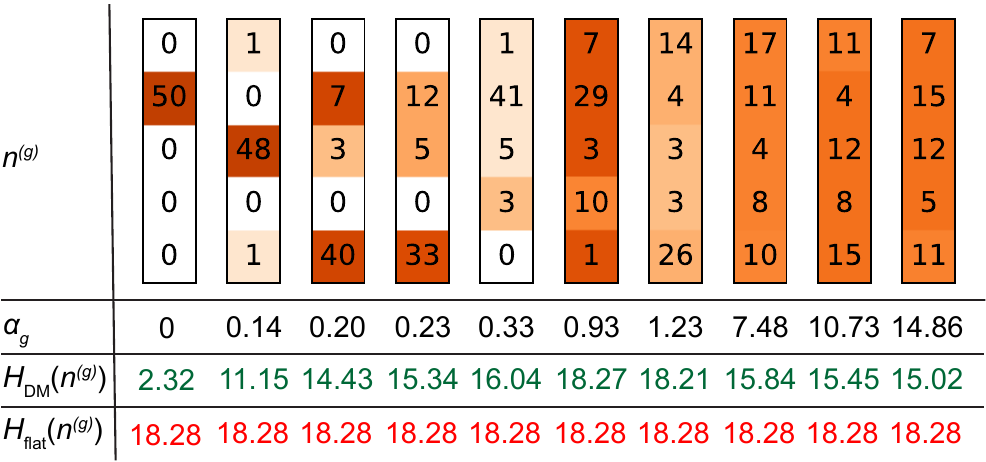}
\caption{Optimal values of~$\alpha_g$ for the transmission of the community sizes~$n^{(g)}$, along with the resulting information cost in bits $H_{\text{DM}}(n^{(g)})$ in our new Dirichlet-multinomial encoding scheme and the corresponding cost $H_{\text{flat}}(n^{(g)})$ in the old, flat encoding.  Note how vectors with more extreme values benefit from smaller values of~$\alpha_g$, while more uniform vectors favor larger~$\alpha_g$.}
\label{fig:DM-vectors}
\end{figure}

We apply this procedure to a selection of example values of~$n^{(g)}$ in Fig.~\ref{fig:DM-vectors}, giving the optimal values of~$\alpha_g$ for each one, along with the resulting values for the entropy.  In each case, as we can see, the Dirichlet-multinomial encoding is more efficient than the conventional flat encoding, sometimes by a wide margin.  In the extreme case where~$n^{(g)}$ has only a single nonzero entry, the optimal value of~$\alpha_g$ is zero and the information cost~is
\begin{align}
  H(n^{(g)}|q_g,\alpha_g = 0) = \log q_g,
\label{eq:HngGalpha0}
\end{align}
whereas the cost to transmit the same $n^{(g)}$ using a flat encoding (equivalent to~$\alpha_g=1$) is considerably steeper:
\begin{align}
  H(n^{(g)}|q_g,\alpha_g = 1) = \log\binom{n + q_g - 1}{q_g - 1} \simeq q_g \log n.
\end{align}

One could argue that to truly make a fair comparison, one should also include the cost to transmit the value of~$\alpha_g$ itself.  As shown in Appendix~\ref{app:alpha-parameter}, however, this cost is small in practice, and moreover cancels completely from the final value of the mutual information, so it is normally safe to ignore it, as we do here.

Although the information saved by using the Dirichlet-multinomial distribution is in some cases a significant fraction of the information needed to transmit~$n^{(g)}$, it is normally quite small next to the information needed to transmit the entire labeling, which is dominated by the cost~$H(g|n^{(g)})$ of sending the labeling itself.  The same is not true, however, when we turn to the conditional entropy~$H(g|c)$, where using the Dirichlet-multinomial distribution can result in large efficiency gains, and this is our primary motivation for taking this approach.  

Recall that the standard model for the conditional information breaks the transmission process into four steps---transmission of~$q_g$,~$n^{(g)}$,~$n^{(gc)}$, and~$g$---which can be represented by the equation
\begin{align}
H_{\text{flat}}(g|c) &= H(q_g) + H(n^{(g)}|q_g) + H(n^{(gc)}|n^{(g)},n^{(c)})
  \nonumber\\
  &\qquad + H(g|c,n^{(gc)}),
\end{align}
with a flat encoding at each step.  In our alternate proposal, we again transmit~$q_g$ using a flat encoding, but then combine the second and third steps to transmit the contingency table all at once, given~$q_g$ and~$n^{(c)}$.  This transmission again leverages a nonuniform encoding to achieve efficiency gains.  The final step of transmitting~$g$ itself remains unchanged.

Our process for transmitting the contingency table involves transmitting one column at a time using the Dirichlet-multinomial distribution.  We use the same value~$\alpha_{g|c}$ for each column, but the columns are otherwise independent.  If we denote column~$s$ by~$n_{\cdot s}^{(gc)}$, then the information cost of this procedure can be written
\begin{align}
H(&n^{(gc)}|n^{(c)}, q_g, \alpha_{g|c}) = \sum_{s=1}^{q_c} H(n_{\cdot s}^{(gc)}|n_s^{(c)},q_g,\alpha_{g|c}) \nonumber\\
  &= \sum_{s=1}^{q_c}\biggl[\log \binom{n_s^{(c)} + q_g \alpha_{g|c} - 1}{q_g \alpha_{g|c} - 1} \nonumber\\
  &\hspace{6em} - \sum_{r=1}^{q_g}\log \binom{n_{rs}^{(gc)} + \alpha_{g|c} - 1}{\alpha_{g|c} - 1}\biggr]. \label{eq:HngcGalphagc}
\end{align}

Consider for instance the special (but not implausible) case where~$g = c$, so that~$n^{(gc)}$ is a diagonal matrix and each column~$n_{\cdot s}^{(gc)}$ has only a single nonzero entry.  Then, as in Eq.~\eqref{eq:HngGalpha0},~$\alpha_{g|c} = 0$ is the optimal choice for transmitting the contingency table and the total information cost is simply
\begin{align}
    H(n^{(gg)}|n^{(g)}, \alpha_{g|c} = 0) &= \sum_{r = 1}^{q_g} H(n_{.r}^{(gg)}|n_r^{(g)},\alpha_{g|c} = 0) \nonumber \\
    &= \sum_{r = 1}^{q_g} \log q_g = q_g \log q_g.
\label{eq:HnggGngalpha0}
\end{align}
In the traditional flat encoding the cost is much greater:
\begin{align}
    H(n^{(gg)}|n^{(g)}) &= H(n^{(g)}|n,q_g) + H(n^{(gc)}|n^{(c)},n^{(g)}) \nonumber\\
    &= \log \binom{n - 1}{q_g - 1} + \log \Omega(n^{(c)},n^{(q)}) \nonumber\\
    &= \Ord(q_g q_c \log n).
\end{align}
This significant improvement also extends to the case where~$g \simeq c$ and the labelings are similar but not identical.  This is precisely the regime in which these measures are typically applied to quantify similarity, so that in realistic settings the new encoding is much preferred over the old one, a conclusion strongly confirmed by the example applications given in Section~\ref{sec:results}.

\begin{figure*}
\centering
\includegraphics[width=\linewidth]{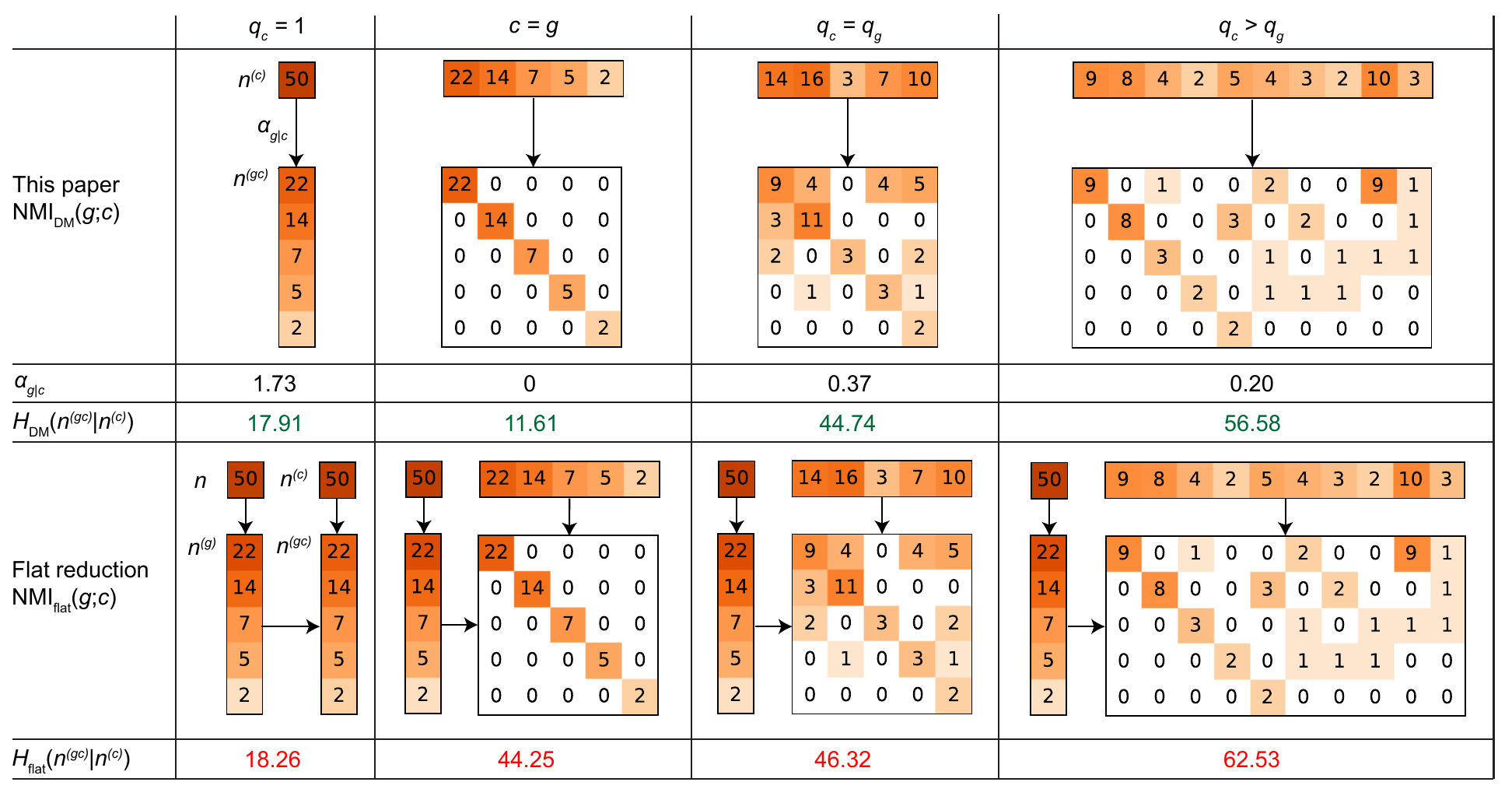}
\caption{Comparison of the information cost of transmitting example contingency tables under the old and new encodings.  The top half of the figure shows the new encoding, along with the values for the optimal Dirichlet-multinomial parameter~$\alpha_{g|c}$ and the resulting information cost~$H_{\text{DM}}(n^{(gc)}|n^{(c)})$.  The bottom half shows the old encoding and associated information cost~$H_\text{flat}(n^{(gc)}|n^{(c)})$.  Note that the new encoding is more efficient in every case, and especially so in the case of equal labelings~$c=g$.}
\label{fig:DM-tables}
\end{figure*}

Figure~\ref{fig:DM-tables} shows the equivalent of Fig.~\ref{fig:DM-vectors} for the transmission of a selection of example contingency tables.  In the case where the candidate labeling places all objects in a single group, so that~$q_c=1$, our proposed scheme is exactly analogous to our method for transmitting the vector~$n^{(g)}$, which implies that the mutual information is~$I_{\text{DM}}(g;c) = 0$ (the DM denoting ``Dirichlet-multinomial'').  This is a desirable property which is also shared by the traditional and reduced mutual informations---a candidate labeling that places all objects in a single group tells us nothing about the ground truth~$g$.  We also note the considerable gulf in efficiency between the two encodings for the case of identical labelings~$g=c$, the second column in Fig.~\ref{fig:DM-tables}, while for labelings that are dissimilar (the final two columns of the figure), the gains of the new encoding are more modest, as we would expect.

Employing our new encoding in the calculation of the conditional information, we now obtain a revised information cost of
\begin{align}
  H_{\text{DM}}(g|c) &= H(q_g) + H(n^{(gc)}|n^{(c)},q_g,\alpha_{g|c})
  \nonumber\\
  &\qquad + H(g|c,n^{(gc)}).
\label{eq:HDMgGc}
\end{align}
(Once again one could arguably also include the fixed cost of transmitting the value of~$\alpha_{g|c}$, but in practice this cost is small and moreover cancels from the final value of the mutual information---see Appendix~\ref{app:alpha-parameter}.)

Putting everything together, we then arrive at our improved mutual information measure 
\begin{align}
I&_{\text{DM}}(g;c) = H_{\text{DM}}(g) - H_{\text{DM}}(g|c) \nonumber\\
    &= I_0(g;c) + H(n^{(g)}|n,q_g,\alpha_{g}) - H(n^{(gc)}|n^{(c)},q_g,\alpha_{g|c}) \nonumber\\
    &= I_0(g;c) \nonumber\\
    &\qquad + \log \binom{n + q_g \alpha_g - 1}{q_g \alpha_g - 1}
    - \sum_{r=1}^{q_g} \log \binom{n_r^{(g)} + \alpha_g - 1}{\alpha_g - 1} \nonumber\\
    &\qquad - \sum_{s = 1}^{q_c} \log \binom{n_s^{(c)} + q_g \alpha_{g|c} - 1}{q_g \alpha_{g|c} - 1} \nonumber\\
    &\qquad + \sum_{r = 1}^{q_g}\sum_{s = 1}^{q_c} \log \binom{n_{rs}^{(gc)} + \alpha_{g|c} - 1}{\alpha_{g|c} + 1}.
\label{eq:ourmeasure}
\end{align}
Besides giving an improved estimate of the mutual information, this formulation has a number of additional advantages over the standard reduced mutual information.  The closed-form expression means that approximations like those used for the number~$\Omega(n^{(g)},n^{(c)})$ of contingency tables in Eq.~\eqref{eq:omega-approx} are unnecessary---the measure can be calculated exactly without approximation.  The measure also has the same advantage over other measures like the adjusted mutual information of Vinh~\etal~\cite{VEB10}, for which the corresponding correction term is calculated using numerically costly Monte Carlo methods.  Another advantage of our proposed measure is that it is possible to prove that~$I_{\text{DM}}(g;c) \le I_{\text{DM}}(g;g)$ for all~$c$, with the exact equality holding only when $g$ and $c$ are identical up to a permutation of labels, an intuitive result that is required for proper normalization, but which has not been shown for the standard reduced mutual information.  We give the proof in Appendix~\ref{app:upper-bound}.

This being said, the encoding we use is not necessarily the last word in calculation of the mutual information.  As discussed at the start of this section, all entropy calculations only give bounds on the true value and it is possible that another encoding exists that could give better bounds.  One could imagine trying an approach analogous to that used for the standard reduced mutual information and constraining not only the column sums of the contingency table but also the row sums, while still using a nonuniform distribution over tables subject to these constraints.  Placing more constraints on the contingency table should reduce the number of tables we need to consider and hence save on transmission costs.  This approach, however, turns out to offer little benefit in practical situations because the gains must be offset against the information needed to transmit the row sums.  It turns out that, in the common case where the candidate and ground-truth labelings are similar to one another, the possible values of the row sums are already tightly restricted, even without placing any explicit constraint on them, so that imposing such a constraint saves little information, while the cost of transmitting the row sums is considerable.  In most cases, therefore, this approach is less efficient than the one we propose.

Equation~\ref{eq:ourmeasure} does still have some shortcomings.  For one thing, it is not fully analytic, since the values of the parameters~$\alpha_g$ and~$\alpha_{g|c}$ must be found by numerical optimization (see Appendix~\ref{app:alpha-parameter}).  Also, because of the asymmetric encoding used to capture the contingency table, in which rows and columns are treated differently, the measure is not symmetric under interchange of~$c$ and~$g$.  For typical applications where one is comparing candidate labelings against a single ground truth this does not matter greatly, since the problem is already inherently asymmetric, but there may be cases where a symmetric measure would be preferred.  Lastly, the encoding we propose is not guaranteed to always perform better than the standard (flat) reduced information.  In particular, if the pairs of labelings~$g$ and~$c$ are truly drawn from the distribution corresponding to the flat encoding scheme, then by definition the flat mutual information will give an optimal encoding and our method cannot do better.  Our broader claim, however, is that in the realistic regime of labelings that have a significant degree of similarity, our new encoding can be expected to perform better than the flat encoding.

\subsection{Normalized mutual information}
So far we have defined various measures of absolute information content, as quantified in bits for example, but such absolute measures can be difficult to interpret.  Are 20 bits of mutual information a little or a lot?  To make sense of the results, they are often expressed in terms of a normalized mutual information (NMI) that represents the information content as a fraction of its maximum possible value~\cite{DDDA05}.  There are various ways to perform the normalization~\cite{MGH11}.  Here we use the form
\begin{align}
\text{NMI}(g;c) = \frac{I(g;c)}{I(g;g)}.
\label{eq:NMI-asym}
\end{align}
Note that this expression is asymmetric in $g$ and $c$---the value is not invariant under their interchange.  Other normalized mutual information measures use a symmetric denominator, such as $\frac12[I(g;g)+I(c;c)]$~\cite{DDDA05,MGH11}, but for reasons given elsewhere~\cite{JKN24} we believe the asymmetric measure to be less biased.

We can define a normalized mutual information for any of the mutual information measures discussed in this paper, including the Dirichlet-multinomial measure.  All of the resulting versions of NMI have the desirable properties of being 1 when $c=g$ and zero when the mutual information is zero.  Thus NMI values approaching 1 generally indicate similar labelings and values near zero indicate dissimilar ones, making this an intuitive measure of similarity.  It is also possible for the NMI to become (slightly) negative for the reduced mutual information measures we consider~\cite{NCY20}.  (This cannot happen with the traditional unreduced mutual information.)  A negative value indicates that the encoding scheme that makes use of $c$ when transmitting $g$ is actually less efficient than simply transmitting $g$ alone.  This, however, happens only when $c$ and $g$ are very dissimilar and hence rarely occurs in practical situations (where we are usually concerned with candidates $c$ that are similar to the ground truth).

For the particular case of the Dirichlet-multinomial mutual information, the NMI has the additional desirable property that it takes the value~1 if and only if $c$~and~$g$ are identical up to a permutation of labels, while for all other $c$ it is less than~1---see Appendix~\ref{app:upper-bound}.  This follows directly from the inequality $I_\text{DM}(g;c) \le I_\text{DM}(g;g)$ mentioned above.  The same is not true of the conventional unreduced NMI, which can be 1 even for very dissimilar labelings (see Section~\ref{sec:reduced-mutual-information}).  It is potentially true, but currently unproven, for the flat reduced mutual information.

\begin{figure*}
\centering
\includegraphics[width=\linewidth]{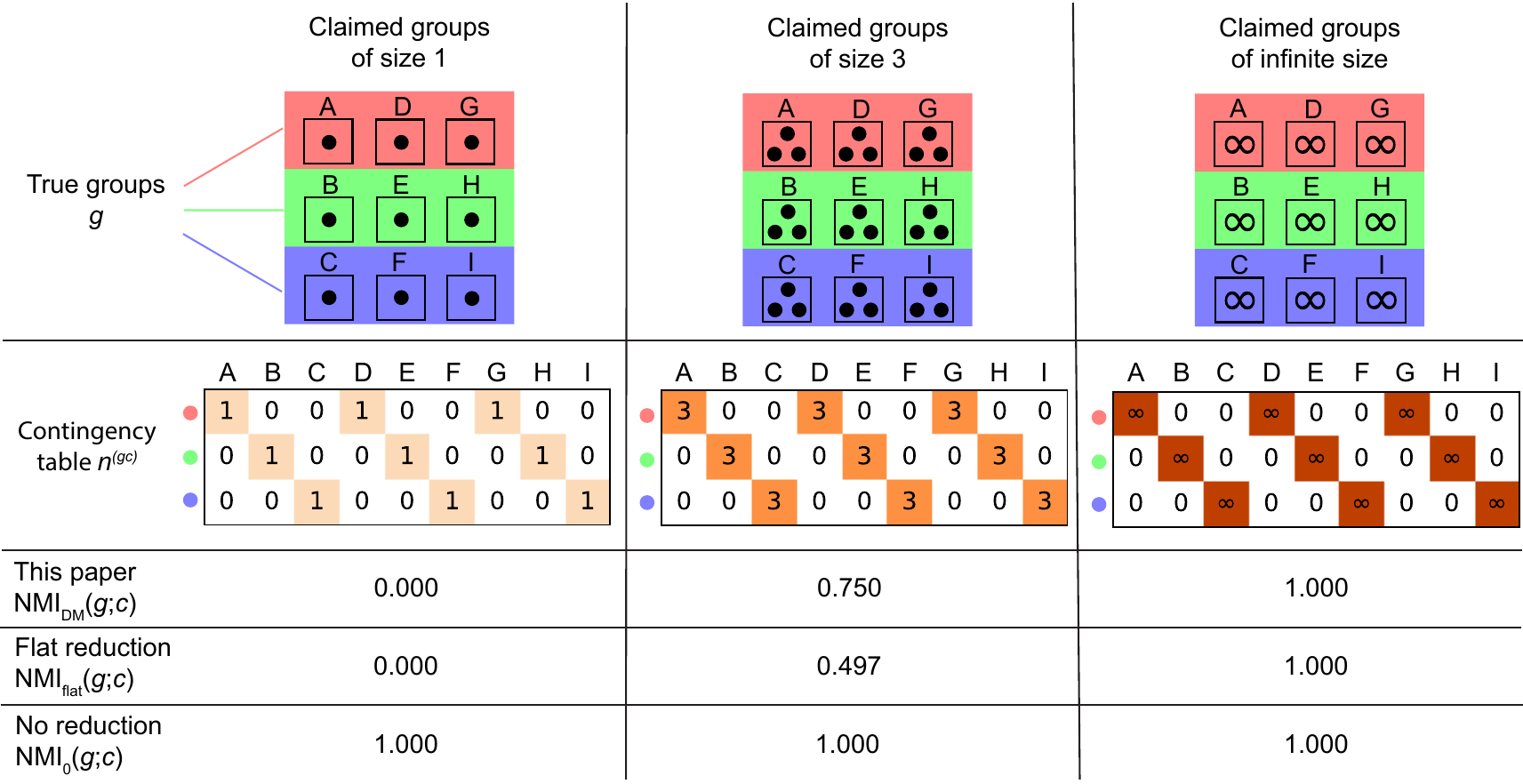}
\caption{Examples of situations where the three normalized mutual information measures considered here can differ substantially.  A set of objects (circular dots) are divided into three ground-truth groups, represented by the horizontal stripes of red, green, and blue.  Three competing candidate divisions of the same objects are represented by the black boxes denoted A to~I.  In each, the ground-truth groups are divided into a set of equally sized subgroups.  But regardless of how many objects are in each subgroup, the unreduced measure~$\text{NMI}_0$ 
returns a maximal score of~1 for all the candidate divisions, while the reduced measures rightfully return lower scores, except in the case where the sizes of the subgroups diverge.  For subgroups of intermediate size, however, such as the groups of size three in the middle column, the Dirichlet-multinomial measure~$\text{NMI}_{\text{DM}}$ of this paper can give a substantially different, and larger score compared to the standard (``flat'') reduced mutual information~$\text{NMI}_{\text{flat}}$.}
\label{fig:tripled-groups}
\end{figure*}

%%%%%%%%%%%%%%%%%%%%%%%%%%%%%%%%%%%%%%%%%%%%%%%%%%%%%%%%%%%%%%%%%%%%%%%%%%%%%%%%%%%%%%%%%%%%%%%%%%%%%%%%%%%%%%%%%%%%%%%%%%%%%%%%%%%
\section{Example applications}
\label{sec:results}
In this section we give a selection of example applications of our proposed measure, demonstrating that it can give significantly different answers from previous measures---different enough to affect scientific conclusions under real-world conditions.

\subsection{Comparison of the proposed measure and the standard reduced mutual information}
In some circumstances the results returned by the measure proposed in this paper can diverge significantly from those given by either the non-reduced mutual information or the standard (``flat'') version of the reduced mutual information.  We have already seen examples for the non-reduced measure: cases in which the candidate labeling~$c$ has many more labels than the ground truth can cause the unreduced measure to badly under\-estimate the true information cost, sometimes maximally so---see Section~\ref{sec:reduced-mutual-information}.

A simple example illustrating the difference between the Dirichlet-multinomial and flat versions of the reduced mutual information is shown in Fig.~\ref{fig:tripled-groups}.  In this example a set of objects, represented by the dots in the figure, are split into three equally sized ground-truth groups.  The candidate labeling~$c$ respects this division but further splits each of the three groups into three subgroups, also of equal size.  This is a special case of the situation mentioned above in which the candidate division has more labels than the ground truth, so it comes as no surprise that the conventional, unreduced mutual information overestimates similarity in this case---indeed it returns the maximal value of~1.

To understand the behavior of our two reduced mutual information measures we consider three special cases.  In the first, shown in the left column of Fig.~\ref{fig:tripled-groups}, each of the subgroups, labeled A to~I, has size~1, meaning that every group in~$c$ has only a single object in it.  We discussed this case previously in Section~\ref{sec:reduced-mutual-information} and argued that the correct mutual information should be zero.  As the figure shows, both versions of the reduced mutual information give this correct result, while the unreduced measure is maximally incorrect.

Next, consider the right column of Fig.~\ref{fig:tripled-groups}, which shows what happens as the total number of objects tends to infinity and the size of the subgroups A to I diverges.  Asymptotically, the candidate~$c$ now gives full information about the ground truth---$g$~is fully specified when both $c$ and the contingency table are known, but the information cost of transmitting the contingency table is a vanishing fraction of the total.  Thus the NMI should be~1 in this case, and again both versions of the reduced mutual information give the right answer.  (In this limit the unreduced measure also gives the right answer.)

But now consider the middle column of the figure, in which subgroups A to I have size~3.  In this case the contingency table, as shown in the figure, is highly non-uniform, and hence is transmitted much more efficiently by the Dirichlet-multinomial encoding than by the flat encoding.  This produces a substantial difference between the values of the two reduced mutual information measures.  The Dirichlet-multinomial measure gives a relatively high value of 0.75, indicating a strong similarity between candidate and ground truth, while the standard flat measure gives a significantly smaller value, less than~0.5.  This is a case where the standard measure has penalized the mutual information too heavily by overestimating the information content of the contingency table, thereby giving a misleading impression that the two labelings are more dissimilar than in fact they are.

\begin{figure}
\centering
\includegraphics[width=8.3cm]{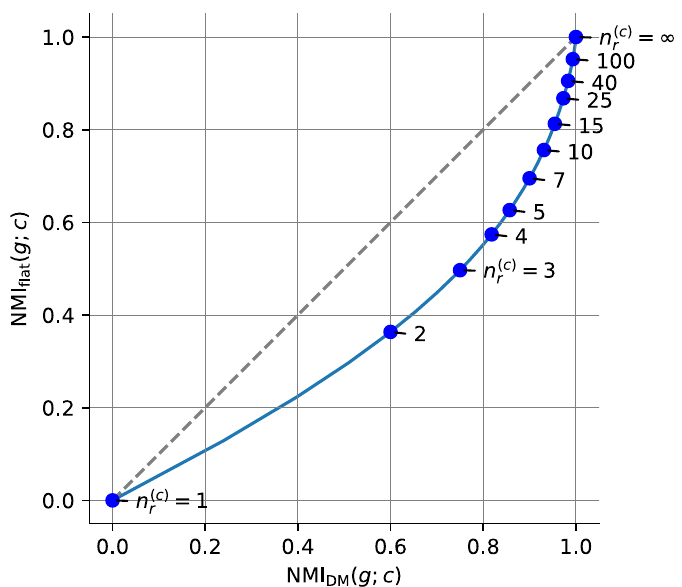}
\caption{Comparison of the two normalized reduced mutual information measures considered here for the example system in Fig.~\ref{fig:tripled-groups}, for various sizes~$n_r^{(c)}$ of the subgroups as denoted by the labels.  In all cases (except $n^{(c)}_r=1$ and~$\infty$) the flat reduced mutual information~$\text{NMI}_{\text{flat}}$ returns a lower value than the Dirichlet-multinomial reduced mutual information~$\text{NMI}_{\text{DM}}$ because it overestimates the information cost of the contingency table.}
\label{fig:tripled-groups-plot}
\end{figure}

Figure~\ref{fig:tripled-groups-plot} shows a plot of the difference between the two reduced measures for this example system with subgroup sizes~$n_r^{(c)}$ ranging all the way from 1 to~$\infty$.  Across the entire range we observe that, apart from the limiting values of $n_r^{(c)}=1$ and~$\infty$, the flat reduced mutual information consistently gives under\-estimates relative to the Dirichlet-multinomial measure.

\begin{figure}
\centering
\includegraphics[width=\linewidth]{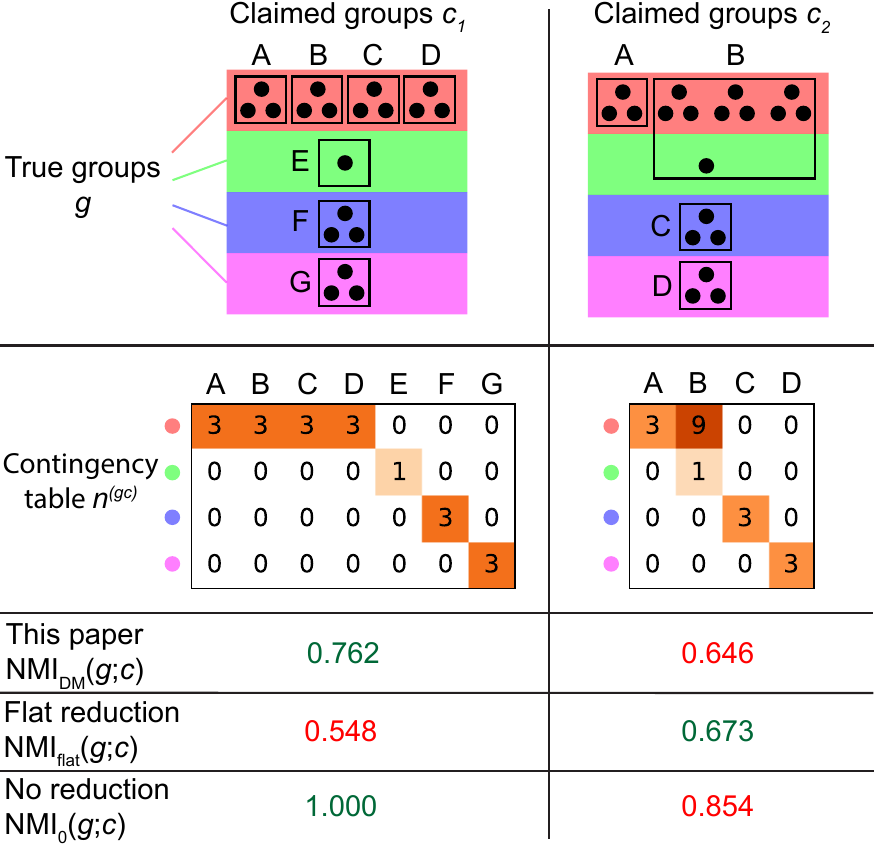}
\caption{Two candidate labelings of the same set of objects.  In this figure, 19~objects are divided among four ground-truth groups, represented by the horizontal stripes of red, green, blue, and magenta, and two candidate labelings~$c_1$ and~$c_2$ are denoted by the boxes labeled A to~G.  The Dirichlet-multinomial measure of this paper favors the left labeling~$c_1$ while the flat reduced mutual information prefers the right one~$c_2$.}
\label{fig:swap-example}
\end{figure}

Figure~\ref{fig:swap-example} shows a different aspect of the two reduced measures.  In this example the ground-truth labeling divides a set of 19 objects into four groups of varying sizes, and we compare outcomes for two proposed candidate labelings, denoted~$c_1$ and~$c_2$.  Labeling~$c_1$ has identified the four groups correctly but has split one of them into a further four subgroups.  As we would expect, the conventional unreduced NMI awards this labeling a maximal score of~1, which is clearly incorrect.  Both reduced measures correctly give a value less than~1, although the values are somewhat different.

Now consider candidate labeling~$c_2$, which erroneously amalgamates the second ground-truth group with part of the first as shown.  Most observers would probably say that this labeling is worse than~$c_1$, but that is not what the standard reduced mutual information reports: the standard measure favors $c_2$ over $c_1$ by a substantial margin.  On the other hand, the Dirichlet-multinomial measure of this paper correctly favors~$c_1$, by a similar margin.

\subsection{Network community detection}
The examples of the previous section are illustrative but anecdotal.  To shed light on the performance of the new measure in a broader context we apply it to the outcomes of a large set of network community detection calculations.  In these tests we use the popular Lancichinetti-Fortunato-Radicchi (LFR) graph model~\cite{LFR08} to generate 100\,000 random networks with known community structure and realistic distributions of node degrees and group sizes.  Then we use six different popular community detection algorithms to generate candidate divisions of these networks, which we compare to the known structure using both the conventional reduced mutual information and the measure proposed here.  Some technical details of the calculations are given in Appendix~\ref{app:LFR}.

\begin{figure}
\centering
\includegraphics[width=\linewidth]{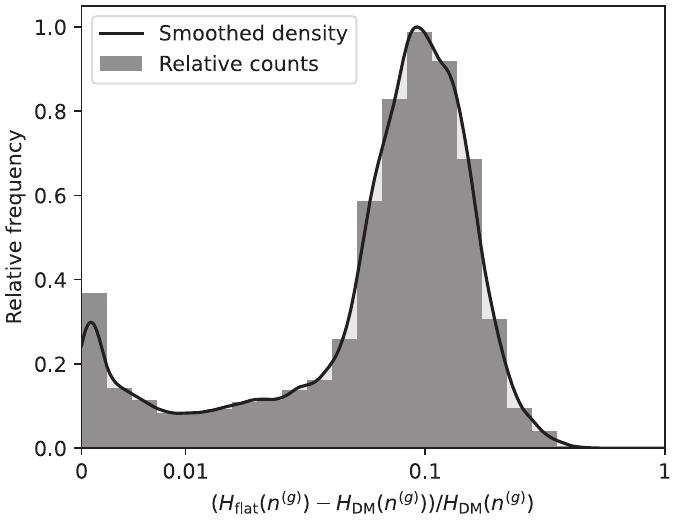}
\caption{Fractional reduction in information cost when transmitting the group sizes~$n^{(g)}$ under the Dirichlet-multinomial encoding versus the flat encoding.  The horizontal axis is linear from 0 to 0.01 and logarithmic above 0.01. The bar chart is a simple histogram of the relative frequencies, while the curve shows the same results smoothed using a quartic kernel density estimator with the same bin width.}
\label{fig:Hng-savings}
\end{figure}

As discussed in Section~\ref{sec:improved-encoding}, our Dirichlet-multinomial approach improves the efficiency of information transmission in two places: in the transmission of the group sizes~$n^{(g)}$ and the transmission of the contingency table~$n^{(gc)}$.  Figure~\ref{fig:Hng-savings} shows the fractional improvement in information cost for the group sizes for each of our test networks.  The gains vary substantially between networks, and some are close to zero, but in a large fraction of cases they reach 10\% or more.

\begin{figure*}
\centering
\includegraphics[width=\linewidth]{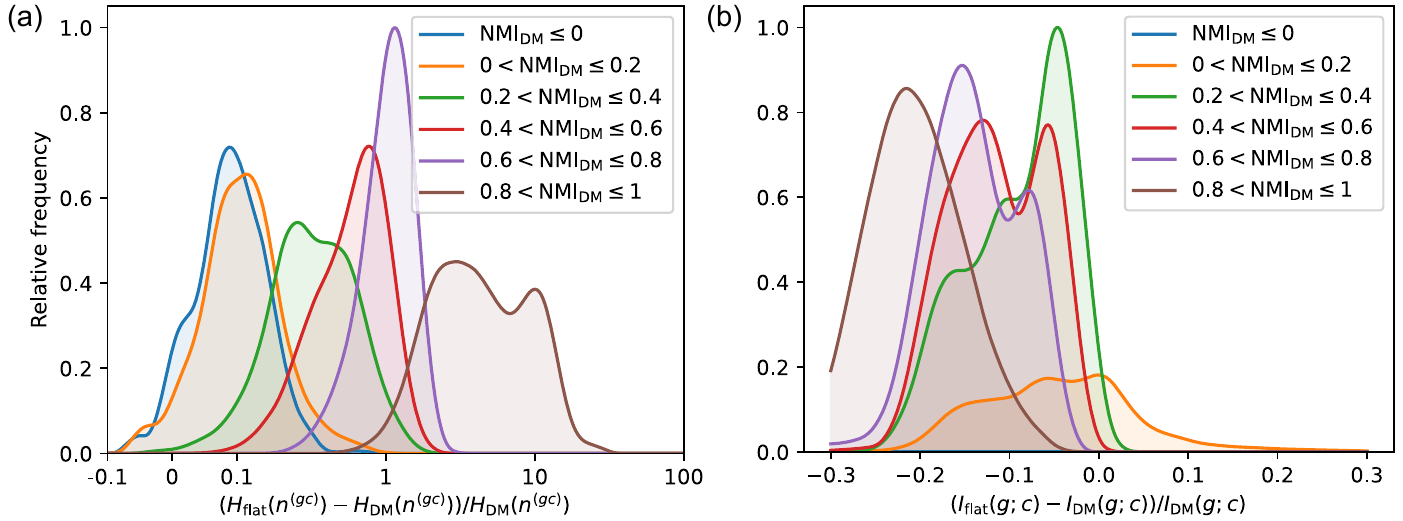}
\caption{(a)~Fractional change in the information cost of transmitting the contingency table~$n^{(gc)}$ using the Dirichlet-multinomial encoding compared to the flat encoding.  The different curves show the distribution of values for different ranges of similarity between the ground-truth and candidate labelings, as measured by the (Dirichlet-multinomial) normalized mutual information.  The horizontal scale is linear between $-0.1$ and 0.1 and logarithmic outside that range.  (b)~Fractional change in the mutual information from the improved encoding of the contingency table and group sizes.  The horizontal axis is linear across the entire range and the different curves again indicate distributions for different ranges of normalized mutual information.  Cases where both mutual informations give a result of 0, for example when $q_c = 1$, have been removed, since they yield a fractional change of $0/0$.}
\label{fig:Hngc-savings}
\end{figure*}

More important, however, are the gains in transmission of the contingency table.  Since the contingency table depends on the candidate~$c$ as well as the ground truth, these gains also depend on the candidate and hence vary between the six different community detection algorithms, but for our purposes here we aggregate the results over algorithms.  Figure~\ref{fig:Hngc-savings}a shows the distribution of the resulting fractional information savings for all networks in a single plot.  The different curves in the plot how the distribution varies as a function of how similar the ground-truth and candidate divisions are, measured using the Dirichlet-multinomial NMI.

Based on these results we observe that when~$g$ and~$c$ are similar ($\text{NMI} > 0.8$, brown curve in the figure) the information gains when transmitting the contingency table are large, up to a factor of ten or more.  This aligns with our observation, discussed in Appendix~\ref{app:upper-bound}, that for~$g\simeq c$ the new encoding scheme is near-optimal, while the flat scheme is very inefficient.  Even in cases where~$g$ and~$c$ are less similar, efficiency gains are often significant, typically above 10\% and as high as 100\% or more.  There are a handful of cases, all occurring when the candidate labeling and ground truth are very dissimilar ($\text{NMI} < 0.2$, blue in the figure), where the new encoding performs slightly worse than the standard one, as discussed in Section~\ref{sec:improved-encoding}.  However, given that mutual information measures are normally applied in cases where the two labelings are significantly similar, the evidence of Fig.~\ref{fig:Hngc-savings} suggests that our new encoding should be preferred, often by a wide margin, in most practical community detection scenarios.

As a result of the changes in both~$H(g)$ and~$H(g|c)$, the value of the mutual information itself can also change significantly.  Figure~\ref{fig:Hngc-savings}b shows the fractional change in the mutual information in our test set, with the different curves again showing the results for different ranges of similarity between the ground truth and the candidate division.  Because the standard encoding usually over\-estimates the conditional information~$H(g|c)$, it tends to underestimate the mutual information~$I(g;c) = H(g) - H(g|c)$, although this bias is offset somewhat by the smaller overestimate of the unconditional entropy~$H(g)$.  On balance, however, the standard encoding significantly underestimates the mutual information in many cases and there are substantial information savings under the new encoding, with the mutual information changing by up to 20\% or more in the common case where the two labelings are similar ($\text{NMI} > 0.8$, brown curve in the figure).

\section{Conclusions}
In this paper we have presented an improved formulation of the mutual information between two labelings of the same set of objects.  Our approach is in the spirit of the recently proposed reduced mutual information: like that measure it addresses the bias towards an excessive number of groups present in traditional measures by taking full account of information costs including particularly the cost of the contingency table.  Where our proposal differs from the standard reduced mutual information is in using a more efficient encoding for the contingency table.  While all information theoretic measures are, in a sense, merely bounds on the true value, our formulation gives significantly tighter bounds in the common regime where the two labelings are similar to one another.

We have demonstrated our proposed measure with a number of examples and performed extensive tests on network community structures generated using the LFR benchmark model.  In the latter context we find that the new encoding does produce considerable savings in information cost and the resulting values for the mutual information differ from the standard reduced mutual information by up to 20\% of the total value under commonly occurring conditions.

Looking ahead, the improved encoding we present for contingency tables could also be used in applications beyond the computation of the mutual information that is the focus of this paper.  In general, the Dirichlet-multinomial distribution that underlies our encoding provides a more informative prior than a standard uniform prior for Bayesian analysis involving contingency tables~\cite{Good76}.  The encoding presented here could thus be used to improve data compression performance of any model that requires the specification of a prior distribution over contingency tables, for example in the methods for clustering discrete data presented in~\cite{Kirkley22,Kirkley23}.

\begin{acknowledgments}
The authors thank Samin Aref for useful comments.  This work was supported in part by the US National Science Foundation under grant DMS--2005899 (MEJN), the Hong Kong Research Grants Council under ECS--27302523 (AK), and by computational resources provided by the Advanced Research Computing initiative at the University of Michigan.  
\end{acknowledgments}

%%%%%%%%%%%%%%%%%%%%%%%%%%%%%%%%%%%%%%%%%%%%%%%%%%%%%%%%%%%%%%%%%%%%%%%%%%%%%%%%%%%%%%%%%%%%%%%%%%%%%%%%%%%%%%%%%%%%%%%%%%%%%%%%%%%
\appendix
\section{Upper bound on the\\normalized mutual information}
\label{app:upper-bound}
In this appendix we show that the value of the normalized mutual information measure proposed in this paper is bounded above by~1:
\begin{align}
\text{NMI}_{\text{DM}}(g;c) = \frac{I_{\text{DM}}(g;c)}{I_{\text{DM}}(g;g)} \leq 1,
\label{eq:NMI-DM-app} 
\end{align}
and moreover that the exact equality is achieved if and only if~$g$ and~$c$ are identical up to a permutation of their labels.  These properties ensure that no candidate can receive a score higher than that of the ground truth itself and enable us to interpret a score of 1 as equality up to permutation.  The conventional (non-reduced) NMI does not have the same properties.  As shown in Fig.~\ref{fig:tripled-groups}, there are possible labelings~$c$ that are substantially different from $g$ but nonetheless give a conventional NMI of~1.  It is possible that the standard (``flat'') reduced mutual information satisfies a bound like~\eqref{eq:NMI-DM-app}, but no such bound has been proven.  It is known to be violated if poor approximations of $\Omega(n^{(g)},n^{(c)})$ are used, so any proof would require an exact expression for $\Omega(n^{(g)},n^{(c)})$ or a sufficiently good estimate.  It is unclear whether current estimates are good enough, although we are not aware of any violations of the relevant inequality when the estimate of Eq.~\eqref{eq:omega-approx} is employed. 

To prove~\eqref{eq:NMI-DM-app} we express the numerator and denominator as
\begin{align}
I_{\text{DM}}(g;c) &= I_0(g;c) + H(n^{(g)}|n,q_g,\alpha_g) \nonumber\\
    &\hspace{5em} - H(n^{(gc)}|n^{(c)},q_g,\alpha_{g|c}), \\
    I_{\text{DM}}(g;g) &= I_0(g;g) + H(n^{(g)}|n,q_g,\alpha_g) \nonumber\\
    &\hspace{5em} - H(n^{(gg)}|n^{(g)},q_g,\alpha_{g|g})\nonumber\\
    &= H_0(g) + H(n^{(g)}|n,q_g,\alpha_g) - q_g\log q_g,
\end{align}
as in Eqs.~\eqref{eq:HnggGngalpha0} and~\eqref{eq:ourmeasure}.  Then our desired bound can be rewritten as
\begin{align}
\log \frac{\prod_s n_s^{(c)}!}{\prod_{rs} n_{rs}^{(gc)}!} + H(n^{(gc)}|n^{(c)},q_g,\alpha_{g|c}) \geq q_g \log q_g. \nonumber\\
\label{eq:qlogq-bound}
\end{align}
The left-hand side of this inequality decreases when the entries of the contingency table decrease.  To demonstrate this we define a table~$\tilde{n}^{(gc)}$ which is identical to the original table~$n^{(gc)}$ except that a single entry is decreased by~1: $\tilde{n}_{rs}^{(gc)} = n_{rs}^{(gc)} - 1$.  With this change the first term in Eq.~\eqref{eq:qlogq-bound} must decrease, since
\begin{equation}
\log \frac{\prod_s n_s^{(c)}!}{\prod_{rs} n_{rs}^{(gc)}!} - \log \frac{\prod_s \tilde{n}_s^{(c)}!}{\prod_{rs} \tilde{n}_{rs}^{(gc)}!} = \log \frac{n_s^{(c)}}{n_{rs}^{(gc)}} \geq 0.
\end{equation}
Similarly, the second term in Eq.~\eqref{eq:qlogq-bound} also decreases if we make the further assumption that
\begin{equation}
n_{rs}^{(gc)} \geq n_{s}^{(c)}/q_g, \qquad n_{rs}^{(gc)} > 1.
\label{eq:nrs-decrement-conditions}
\end{equation} 
Under these conditions we can bound the change in the second term by 
\begin{align}
H(&n^{(gc)}|n^{(c)}, q_g, \alpha_{g|c}) - H(\tilde{n}^{(gc)}|\tilde{n}^{(c)}, q_g, \alpha_{g|c}) \nonumber\\
  &= \log \binom{n_s^{(c)} + q_g \alpha_{g|c} - 1}{q_g \alpha_{g|c} - 1}  - \log \binom{\tilde{n}_s^{(c)} + q_g \alpha_{g|c} - 1}{q_g \alpha_{g|c} - 1}\nonumber\\
  &\quad -\log \binom{n_{rs}^{(gc)} + \alpha_{g|c} - 1}{\alpha_{g|c} - 1} + \log \binom{\tilde{n}_{rs}^{(gc)} + \alpha_{g|c} - 1}{\alpha_{g|c} - 1}\nonumber\\
  &= \log \frac{n_{s}^{(c)}+q_g\alpha_{g|c} - 1}{n_{s}^{(c)}} - \log \frac{n_{rs}^{(gc)}+\alpha_{g|c} - 1}{n_{rs}^{(gc)}}\nonumber\\
  &\geq \log \frac{n_{s}^{(c)}+q_g\alpha_{g|c} - q_g}{n_{s}^{(c)}} - \log \frac{n_{rs}^{(gc)}+\alpha_{g|c} - 1}{n_{rs}^{(gc)}} \nonumber\\
  &= \log \frac{n_{s}^{(c)}/q_g+\alpha_{g|c} - 1}{n_{s}^{(c)}/q_g} - \log \frac{n_{rs}^{(gc)}+\alpha_{g|c} - 1}{n_{rs}^{(gc)}} \nonumber\\
  &\geq 0, \nonumber\\
\end{align}
where in the last step we have made use of~\eqref{eq:nrs-decrement-conditions} and the the fact that $\log((x + \alpha - 1)/x)$ is monotonically decreasing in $x$ for all~$x>0$.

Now we observe that if there is any entry of $n^{(gc)}$ such that $n_{rs}^{(gc)}>1$, then there must be an entry~$n_{rs}^{(gc)} \ge n_{s}^{(c)}/q_g$, i.e.,~it is greater than or equal to the average for its column.  We apply this observation repeatedly to decrement each non-zero entry of the table to 1 until $\tilde{n}_{rs}^{(gc)} = \text{min}\bigl(n_{rs}^{(gc)},1\bigr)$, while at the same time ensuring that
\begin{align}
\log &\frac{\prod_s n_s^{(c)}!}{\prod_{rs} n_{rs}^{(gc)}!} + H(n^{(gc)}|n^{(c)},q_g,\alpha_{g|c}) \nonumber\\
    &\geq \log \frac{\prod_s \tilde{n}_s^{(c)}!}{\prod_{rs} \tilde{n}_{rs}^{(gc)}!} + H(\tilde{n}^{(gc)}|\tilde{n}^{(c)},q_g,\alpha_{g|c}).
\end{align}
This reduces the problem of showing the general inequality~\eqref{eq:qlogq-bound} to proving it for tables $\tilde{n}$ whose entries are 0 or 1 only, which we can do as follows:
\begin{subequations}
\begin{align}
\log &\frac{\prod_s \tilde{n}_s^{(c)}!}{\prod_{rs} \tilde{n}_{rs}^{(gc)}!} + H(\tilde{n}^{(gc)}|\tilde{n}^{(c)},q_g,\alpha_{g|c})\nonumber\\
    &= \sum_s\biggl[\log\tilde{n}_s^{(c)}! + \log \binom{\tilde{n}_s^{(c)} + q_g \alpha_{g|c} - 1}{q_g \alpha_{g|c} - 1} \nonumber\\
    &\hspace{4em} - \sum_r \log \binom{\tilde{n}_{rs}^{(gc)} + \alpha_{g|c} - 1}{\alpha_{g|c} - 1}\biggr]\nonumber\\
    &= \sum_s\bigl[\log (\tilde{n}_s^{(c)} + q_g \alpha_{g|c} - 1)!- \log(q_g \alpha_{g|c} - 1)! \nonumber\\
    &\hspace{4em} - \tilde{n}_s^{(c)} \log \alpha_{g|c}\bigr]\nonumber\\
\label{eq:saturate-inequality-1} 
    &\geq \sum_s \left[\sum_{t=0}^{\tilde{n}_s^{(c)} - 1} \log(q_g \alpha_{g|c} + t) - \tilde{n}_s^{(c)} \log \alpha_{g|c} \right] \\
\label{eq:saturate-inequality-2}
    &\geq \sum_s  \tilde{n}_s^{(c)}\bigl[\log(q_g \alpha_{g|c})-  \log \alpha_{g|c}\bigr] \\
    &\geq \sum_s  \tilde{n}_s^{(c)}\log q_g \geq q_g\log q_g,
\label{eq:saturate-inequality-3}
\end{align}
\end{subequations}
where in the final step we have made use of the fact that each of the~$q_g$ groups must contain at least one object, so there must be at least~$q_g$ nonzero entries in~$n^{(gc)}$ and hence also in~$\tilde{n}^{(gc)}$.  We also observe that the inequalities~\eqref{eq:saturate-inequality-1}-\eqref{eq:saturate-inequality-3} are saturated only when~$\tilde{n}_s^{(c)} = 1$ for all~$s$ and~$q_c = q_g$. These conditions together imply that the contingency table~$n^{(gc)}$ must be diagonal, and hence that the labelings~$g$ and~$c$ are equivalent up to a permutation of their labels.  The reverse conclusion, that $\text{RMI}_{\text{DM}}(g;c) = 1$ when $g$ and $c$ are equivalent up to a permutation, also follows since our measure is invariant under label permutations.

Finally, we note that if we instead normalize the reduced mutual information symmetrically according to
\begin{align}
\text{NMI}_{\text{DM}}^{(S)}(g;c) = \frac{I_{\text{DM}}(g;c) + I_{\text{DM}}(c;g)}{I_{\text{DM}}(g;g) + I_{\text{DM}}(c;c)},
\end{align}
then the results of this section also ensure that~$\text{NMI}_{\text{DM}}^{(S)}(g;c) \leq 1$ and that this bound is saturated only for~$g$ and~$c$ equivalent up to a permutation.  This symmetric normalization may be more appropriate when comparing two labelings neither of which can be considered a ground truth.

%%%%%%%%%%%%%%%%%%%%%%%%%%%%%%%%%
\section{Clustering and permutation invariance}
\label{app:permutations}
In this paper we have focused on the comparison of different labelings of a set of objects, but the most common applications of the mutual information are actually to the comparison of \emph{clusterings}, i.e.,~partitions of objects into some number~$q_g$ of (unlabeled) groups.  One can easily represent a clustering by arbitrarily assigning integer labels $1\ldots q_g$ to the groups and then recording the label of the group to which each object belongs, but the mapping from clusterings to labelings is not unique: here are $q_g!$ permutations of the labels that correspond to the same clustering.  This means that the information cost of transmitting a labeling, as discussed in this paper, is larger than the information cost of transmitting a clustering.  In the most extreme case, suppose that we want to transmit the unique clustering of~$n$ objects into~$n$ distinct groups, with a single object in each group.  There are $n!$ possible labelings that represent this clustering, so the information cost to transmit any one of them is $H_0(g) = \log n!$ as in Eq.~\eqref{eq:H0g}. Yet there is only a single clustering that places each object in its own group, so in principle the information cost should be $\log 1 = 0$.  Thus the label-based approach grossly over\-estimates the true information cost in this case.  As we argue in this appendix, however, the amount of the overestimate is a constant that plays no role in typical applications, and cancels completely from the mutual information itself, so in practice the measures described in this paper give correct and useful answers as is. 

What is the actual information content of a clustering, not just of the labeling that represents it?  To answer this question we adopt a notation that directly describes clusterings rather than labelings.  For a given labeling~$g$ with~$q_g$ labels we define the equivalence class~$\tilde{g}$ to be the set of all~$q_g!$ variants of $g$ obtained by permutations of the label values, including the original permutation $g$ itself.  By combining all these permutations into a single object, the equivalence class directly represents the clustering of which labeling~$g$ is a manifestation.  With this definition we can adapt the encoding schemes for labelings described in this paper to give encoding schemes for clusterings. 

Any encoding of labelings effectively defines a probability distribution over all labelings via~$P(g) = e^{-H(g)}$. Since the schemes of this paper are all invariant under the~$q_g!$ possible permutations of the labels, we can easily sum up the resulting probability weight over all labelings that represent a given clustering to find the induced probability distribution over clusterings:
\begin{align}
P(\tilde{g}) = \sum_{q \in \tilde{g}} P(g) = q_g! \,P(g).
\end{align}
Under this distribution the cost to directly transmit the clustering $\tilde{g}$ independent of its label assignment is \begin{align}
    H(\tilde{g}) = - \log P(\tilde{g}) = H(g) - \log q_g! \label{eq:Htildeg}
\end{align}
Thus, if we could find a way to transmit only the clustering we would realize an information savings of~$\log q_g!$ compared with the transmission of an arbitrary labeling.

A practical way to achieve this is simply to agree upon a single unique labeling that will represent each possible clustering.  Only these agreed labelings will be transmitted and no others.  By definition this reduces the number of possible labelings by a factor of $q_g!$ and hence reduces the information by $\log q_g!$, as above.

To give an explicit example of such an encoding, we could stipulate that every labeling must have the following two properties:
\begin{enumerate}
\item Groups are labeled in order of increasing size, so that group~1 is the smallest and group~$q_g$ is the largest.
\item If two groups have the same size, the tie is broken by giving the smaller group label to the group that appears first in the ordered list of all objects.
\end{enumerate}
For every clustering there is only one labeling that satisfies these rules, and any labeling that does not satisfy them can easily be converted into one that does.  For example,
$g = 33132112$ becomes $22321331$.

If enforcement of the above rules is denoted by~$R$, the information content of a clustering is
\begin{align}
    H(\tilde{g}) = H(g|R),
\end{align}
and with these definitions we can now explicitly calculate the information needed to transmit a clustering.  As before, we transmit the clustering in three steps.  In the first step we transmit the number of groups~$q_g$.  The fact that a labeling respects the rules~$R$ has no effect on~$q_g$, so the information required for this step is unchanged from before: $H(q_g|R) = H(q_g)$.

In the second step we transmit the group sizes~$n^{(g)}$, and here there is a change because rule~1 above implies that the group sizes must appear in non-decreasing order, and hence the possible values of $n^{(g)}$ are drawn only from the set of such non-decreasing candidates, a subset of the $\binom{n - 1}{q_g - 1}$ possible vectors that sum to~$n$.  We further note that not all of these non-decreasing vectors will occur with equal frequency.  The number of ways one such vector can occur in our transmission process is equal to the number of unique starting vectors that can be permuted into the given non-decreasing form. If we define the multiplicity of the group sizes as
\begin{align}
    M_t = \big|\{r|n_r^{(g)} = t\}\big|, \qquad t = 1\dots q_g,
\end{align}
then there are $q_g!/\prod_t M_t!$ such permutations.  So the probability that any individual one will occur is $(q_g!/\prod_t M_t!)/{n-1\choose q_g-1}$ and the information cost to transmit $n^{(g)}$ is minus the log of this probability:
\begin{align}
    H(n^{(g)}|q_g,R) = \log \left[\binom{n-1}{q_g - 1} \frac{\prod_t M_t!}{q_g!}\right].
\end{align}

In the third step of the transmission process we transmit the labeling itself, and here too the information cost is modified because of our rules.  Whenever two groups of the same size are present, we know that the group appearing first must have the smaller group label because of rule~2 above and hence we need only consider labelings that satisfy this requirement.  This leaves only a fraction $1/\prod_t M_t!$ of the original $n!/\prod_r n_r^{(g)}!$ labelings, giving an information cost of
\begin{align}
H(g|n^{(g)},R) = \log \frac{n!}{\prod_r n_r^{(g)}!\prod_t M_t!}.
\end{align}
Combining these terms, the total information cost to transmit the clustering is
\begin{align}
H(\tilde{g}) &= H(g|R) \nonumber\\
    &= H(q_g|R) + H(n^{(g)}|q_g,R) + H(g|n^{(g)},R) \nonumber\\
    &= H(q_g) + H(n^{(g)}) + \log \prod_t M_t! - \log q_g! \nonumber\\
    &\qquad + H(g|n^{(g)}) - \log \prod_t M_t! \nonumber\\
    &= H(g) - \log q_g!
\end{align}
as expected.

Taking, for instance, our earlier example in which there are $n$ groups of one object each, all groups have the same size, so by rule~2 above they are simply labeled in order of their appearance $123\ldots n$.  This is the unique valid labeling with this set of group sizes, so setting $q_g=n$, the information cost is correctly given as $H(g) - \log q_g! = \log n! - \log n! = 0$.

The same discounted information cost also applies to the conditional entropy.  Suppose we are given a candidate clustering denoted by equivalence class~$\tilde{c}$ and represented as above by a unique labeling~$c$ within that class, such as the one obeying the rules~$R$.  Since our encoding schemes are invariant under label permutations, all labelings in~$\tilde{c}$ are equally informative, including the one~$c$ that we have selected, and hence
\begin{align}
    H(\tilde{g}|\tilde{c}) = H(\tilde{g}|c).
\end{align} 
Using the same argument as before, the conditional information cost of the clustering is then given by
\begin{align}
    H(\tilde{g}|c) = H(g|c) - \log q_g!
\end{align}
and hence the mutual information between two \emph{clusterings} is given by
\begin{align}
    I(\tilde{g};\tilde{c}) &= H(\tilde{g}) - H(\tilde{g}|\tilde{c}) \nonumber\\
    &= H(g) - \log q_g! - \bigl[ H(g|c) - \log q_g! \bigr] \nonumber\\
    &= H(g) - H(g|c) = I(g;c).
\end{align}
Thus, the mutual information between clusterings is the same as between any corresponding pair of labelings.  In practice, this means that we never need to consider mutual information measures between clusterings: calculating the mutual information between labelings, as described in this paper, is more straightforward and will give the same result.

Using this clustering perspective we can also show that the encoding we propose in this paper is near optimal in the important case where $c = g$.  All the encoding schemes we consider are invariant under label permutations, which implies that
\begin{equation}
H(g|g) = H(g|\tilde{g}) = H(\tilde{g}|\tilde{g}) + \log q_g! \ge \log q_g!
\label{eq:Hggbound}
\end{equation}
From Eqs.~\eqref{eq:HnggGngalpha0} and~\eqref{eq:HDMgGc} our Dirichlet-multinomial encoding has cost
\begin{align}
H_{\text{DM}}(g|g) &= H(q_g) + H(n^{(gg)}|n^{(g)},\alpha_{g|g}) + H(g|c,n^{(gg)}) \nonumber\\
    &= \log n + q_g \log q_g.
\label{eq:Hdmgg}
\end{align}
If we accept the cost $\log n$ of transmitting the number of groups~$q_g$ as a necessary price of doing business, this value for $H_{\text{DM}}(g|g)$ very nearly saturates the bound in Eq.~\eqref{eq:Hggbound}, since the gap between $q_g \log q_g$ and $\log q_g!$ is only of order $\Ord(q_g)$.  By contrast, the flat encoding is far from saturating the bound in this case, explaining its poorer performance in the important regime where $c\simeq g$.  Equation~\eqref{eq:Hdmgg} also helps explain a point made in Section~\ref{sec:improved-encoding}, that it is rarely beneficial to constrain both the row and column sums of the contingency table, since the Dirichlet-multinomial encoding is already near-optimal while constraining only the columns.

\begin{figure*}
\centering
\includegraphics[width=\linewidth]{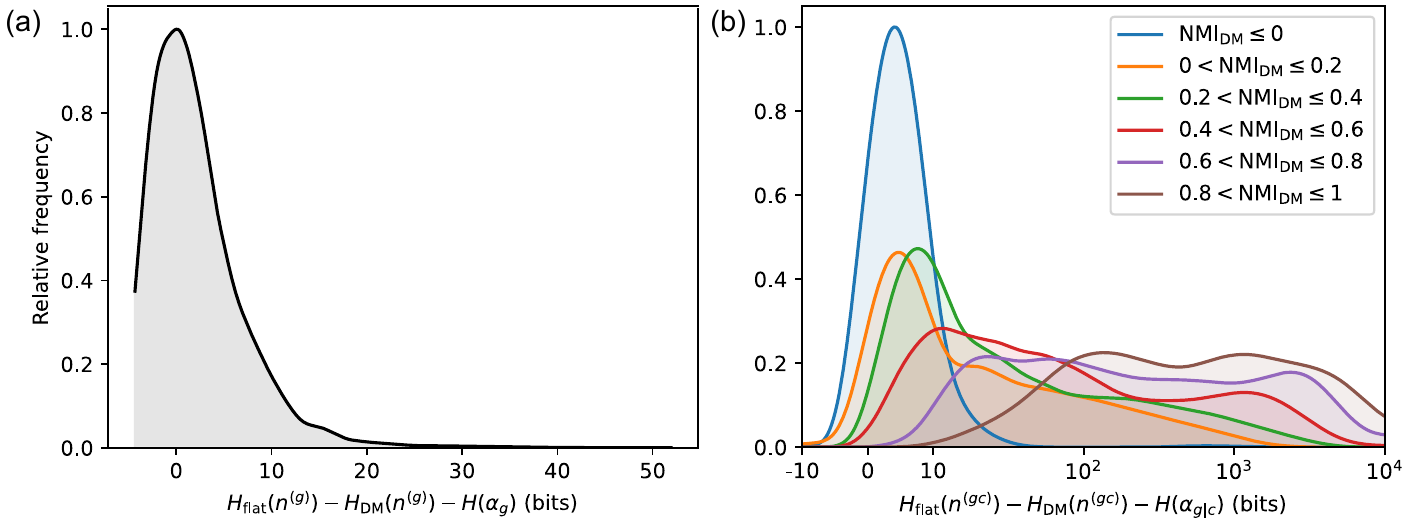}
\caption{(a)~The absolute change in the information cost of transmitting the vector of true group sizes~$n^{(g)}$ between the standard flat encoding and the optimized Dirichlet-multinomial encoding. In contrast with Fig.~\ref{fig:Hng-savings}, the information cost $H(\alpha_g) = 4$ bits for transmitting the Dirichlet-multinomial parameter is included in this comparison.  (b)~The absolute change in the information cost of transmitting the contingency table~$n^{(gc)}$ between the standard flat encoding and the Dirichlet-multinomial encoding (including $H(\alpha_{g|c})$).  The different curves show the distribution for different levels of similarity between the ground-truth and candidate labelings, as measured by the normalized mutual information. The horizontal scale is linear between $-10$ and 10 and logarithmic outside that range.  In both panels the densities of cases are transformed and smoothed as in Figure~\ref{fig:Hng-savings}.}
\label{fig:alpha-cost}
\end{figure*}

%%%%%%%%%%%%%%%%%%%%%%%%%%%%%%%%%
\section{Choosing and transmitting the\\value of the Dirichlet-multinomial parameter}
\label{app:alpha-parameter}
In computing the information costs~$H(n^{(g)}|q_g,\alpha_g)$ and~$H(n^{(gc)}|n^{(c)}, q_g, \alpha_{g|c})$ that appear in Eqs.~\eqref{eq:HngGalphag} and~\eqref{eq:HngcGalphagc}, we have used the values of the Dirichlet-multinomial parameters $\alpha_g$ and~$\alpha_{g|c}$ that minimize those costs.  These values were found by numerical optimization, using golden-ratio search in the space of $\log\alpha$ with a starting bracket of $\alpha\in[10^{-3},10^3]$.

In Figs.~\ref{fig:Hng-savings} and~\ref{fig:Hngc-savings} we compared the information costs of transmitting the group sizes~$n^{(g)}$ and the contingency table~$n^{(gc)}$ within the Dirichlet-multinomial encoding scheme and the standard (flat) encoding, but we neglected the cost of sending the value of~$\alpha$, which arguably means the comparison is not entirely fair.  Assigning a cost to the transmission of $\alpha$ is somewhat delicate, since it is a continuous-valued parameter with a potentially infinite number of decimal digits, and hence its complete transmission would require an infinite amount of information.  In practice, however, high accuracy is not needed to get most of the benefit of the Dirichlet-multinomial approach and we can use a small number of bits to transmit a value chosen from a finite set of possibilities without losing much.  For example, by using four bits of information we can transmit a value chosen from the sixteen possibilities $\alpha \in \{10^{-2},10^{-1.75},10^{-1.5},\ldots,10^{1.5},10^{1.75}\}$.  In Figure~\ref{fig:alpha-cost} we show the resulting difference in information cost between the Dirichlet-multinomial and flat encodings when this additional small cost is taken into account.  As panel~(a) shows, the cost of transmitting $\alpha_g$ does have a noticeable effect on the (already small) information to transmit $n^{(g)}$, the flat encoding now being favored in a number of cases, but this is usually not an issue, since the information cost of $n^{(g)}$ is not a large part of the total in most practical situations.  As panel~(b) shows, we retain the significant gains in the transmission of the contingency table under the Dirichlet-multinomial scheme, even allowing for the cost of transmitting~$\alpha_{g|c}$, especially in the common regime where $c\simeq g$.

Moreover, these concerns will not impact our final mutual information score at all if the same method is used to transmit both~$\alpha_g$ and~$\alpha_{g|c}$.  Any costs that we include will cancel in the expression for the mutual information because
\begin{align}
    I&_{\text{DM}}(g;c) = I_0(g;c) + H(n^{(g)}|n,q_g,\alpha_{g}) + H(\alpha_g) \nonumber\\
    &\hspace{6em} - \bigl[H(n^{(gc)}|n^{(c)},q_g,\alpha_{g|c}) + H(\alpha_{g|c})\bigr] \nonumber\\
    &= I_0(g;c) + H(n^{(g)}|n,q_g,\alpha_{g}) - H(n^{(gc)}|n^{(c)},q_g,\alpha_{g|c}).
\end{align}
In practice, therefore, the cost of transmitting $\alpha$ plays no role in our calculation of the mutual information.

%%%%%%%%%%%%%%%%%%%%%%%%%%%%%%%%%%%%%%%%%%%%%%%%%%%%%%
\section{Benchmark generation}
In this appendix we briefly describe the generation of benchmark networks and the community detection algorithms used in our network clustering tests.

\subsection{LFR network generation}
\label{app:LFR}
The networks we use for benchmarking are generated using the LFR model described in~\cite{LFR08}, which creates networks with relatively realistic features by the following procedure.
\begin{enumerate}
	\item \textbf{Fix the number of nodes~$n$ and mixing parameter~$\mu$}.  In our examples we use node counts in the range~$n \in [200,51200]$.  The parameter~$\mu$ controls the relative number of edges within and between communities.  For small~$\mu$ there are many more edges within communities than between them, which makes the communities relatively easy to detect.  But as~$\mu$ increases there are more edges between communities and detection becomes more difficult.  Our examples span values of~$\mu$ in the range~$[0.2,0.8]$.
	\item \textbf{Draw a degree sequence} from a power-law distribution with exponent~$\tau_1$.  Many networks have power-law degree distributions, typically with exponents between 2 and~3~\cite{Caldarelli07}, and the LFR model exclusively uses power-law distributions.  We use~$\tau_1 = 2.5$, with average degree~$\langle k \rangle = 20$ and a maximum degree that scales with graph size as~$k_{\text{max}} = n/10$.
	\item \textbf{Draw a set of community sizes} from a power-law distribution with exponent~$\tau_2$.  Many networks also have community sizes that approximately follow a power law, with typical exponents in the range from 1 to~2~\cite{Guimera03,CNM04,PDFV05,LFR08}.  We use~$\tau_2 = 1.5$ and a minimum community size of~$s_{\text{min}} = 20$ in all cases, while the maximum community size is set to~$s_{\text{max}} = \max(n/10,100)$.  Empirically, our results are not very sensitive to the choices of degree and community size distributions.
	\item \textbf{Assign each node to a community} at random while ensuring that the community chosen is always large enough to support the added node's intra-community degree, given by~$(1 - \mu) k$ where~$k$ is the total degree.  
    \item \textbf{Rewire the edges} attached to each node while preserving the node degrees so that the fraction of edges connected to each node running outside its community is approximately~$\mu$.
\end{enumerate}
The parameter values above are similar to those used for instance in~\cite{YAT16}. 

\subsection{Community detection algorithms}
\label{app:algorithms}
We perform community detection on the LFR networks using six well-known algorithms to generate realistic pairs~$(g,c)$ of ground truths and candidates as follows. (We use the implementations found in the \verb|igraph| library~\cite{CN06}, except for the inference method, for which we use the \verb|graph-tool| library~\cite{P14}.)
\begin{enumerate}
\item \textbf{InfoMap:} InfoMap is an information theoretic community detection method that defines a compression algorithm for encoding a random walk on a network based on the communities that the walk passes through~\cite{RB08}.  Different community labelings yield different compression efficiencies, as quantified by the so-called map equation, and the labeling with the highest efficiency is considered the best community division.  
\item \textbf{Modularity maximization:} Modularity is a quality function for community divisions equal to the fraction of edges within communities minus the expected such fraction in a randomized version of the network.  Modularity maximization algorithms work by searching for the division of the network that maximizes this modularity.  Exact maximization is NP-hard and computationally intractable in most practical situations, but the modularity can be approximately maximized using various methods such as the Louvain and Leiden algorithms~\cite{BGLL08, TWV19}, spectral methods~\cite{Newman06b}, and simulated annealing~\cite{GSA04,MAD05,RB06a}. 
\item \textbf{Modularity with enhanced resolution:} Standard modularity maximization is known to suffer from a ``resolution limit''---it cannot detect communities smaller than a certain threshold size~\cite{FB07}.  This can be remedied by generalizing the modularity to include a resolution parameter~$\gamma$ such that higher values of~$\gamma$ push the algorithm towards smaller communities~\cite{RB06a}.  Standard modularity maximization corresponds to~$\gamma=1$, but for comparison we also conduct tests with~$\gamma=10$ using the Leiden algorithm.
\item \textbf{Statistical inference:} Community detection can also be formulated as a statistical inference problem.  In this approach one assumes the network to have been generated from a randomized model in which the probabilities of edges depend on the group membership of the nodes at their ends.  Then finding the communities in a given network becomes a question of fitting the model to the network to find the best set of group assignments.  Here we fit the so-called degree-corrected stochastic block model~\cite{KN11a} to our LFR networks using a Bayesian method~\cite{Peixoto17}.
\item \textbf{Walktrap:} Walktrap is an agglomerative algorithm in which initially separate nodes are combined into progressively larger communities in order from strongest to weakest connections, where strength is defined in terms of the time for a random walk to reach one node from another~\cite{PL05}.
\item \textbf{Labelprop:} The label propagation or ``labelprop'' algorithm likewise initially places every node in its own community, then it iteratively updates the labels of randomly chosen nodes by majority vote among their network neighbors~\cite{RRS07}.
\end{enumerate}

% \bibliographystyle{numeric}
% \bibliography{journals,references}

\end{document}